\documentstyle[preprint,aps,epsfig]{revtex}

\begin{document}
\draft
\title{Chiral Lagrangian \\from gauge invariant,
nonlocal, dynamical quark model}

\author{Hua Yang$^{a,b}$, Qing Wang$^{a,c}$, Qin Lu$^a$}

\address{$^a$Department of Physics, Tsinghua University, Beijing 100084, China
\footnote{Mailing address}\\
$^b$Institute of Electronic Technology, Information Engineering University,
Zhengzhou 450004\\
$^c$Institute of Theoretical Physics, Academia Sinica, Beijing 100080, China}

%
\maketitle
\begin{abstract}
Parameters of Gasser Leutwyler chiral Lagrangian are proved
saturated by dynamical quark self energy $\Sigma(k^2)$ in a
 gauge invariant, nonlocal, dynamical quark model.
\end{abstract}

\bigskip


\vspace{1cm}
Much of low-energy QCD can be encoded into a series parameters appearing in
 a chiral Lagrangian, expanded to some finite order of
 low energy expansion. Attempts have been made to understand these
 parameters: it is shown that
 low lying vector mesons will saturate the parameters
 \cite{vectormeson}.
  To go beyond phenomenological level, the anomaly contribution was taken as
 the main source of the parameters \cite{anomalyderive}, we call this type investigation the anomaly approach,it leads result
  \begin{eqnarray}
 8L_1=4L_2=-2L_3=24L_7=-8L_8=L_9=-2L_{10}=\frac{N_c}{48\pi^2}
 \label{anomalyresult}
 \end{eqnarray}
 which are  close to experiment result except $L_7$ and $L_8$ which have wrong signs. The deficiency of this calculation
lies in its independence of interaction: if we switch off the strong
interaction and discuss a system of free quark field with external sources,
the anomaly calculation can still be performed without any change. Then it
seems that (\ref{anomalyresult}) is not due to strong interaction among
quarks and gluons,
 but rather an artificial result. Another type research, we call it
dynamical approach, mainly consider the dynamical effect \cite{holdom},
in which the main source of the parameters is from dynamical quark self energy
$\Sigma(k^2)$. This approach
 has advantage of maintaining chiral symmetry and momentum dependence of
 dynamical quark mass, in the mean time avoiding introduce in the theory the
 hard constituent quark mass to cause wrong bad ultraviolet behavior of the theory. But it does not explain why it can offer the better numerical result
 (without wrong sign problem for $L_7$ and $L_8$) than anomaly approach. In
  fact, anomaly contribution and dynamical  quark self energy contribution are
   two independent sources, if the anomaly contribution play role, according to
 (\ref{anomalyresult}), it will be dominant at all parameters and then there is no room left for $\Sigma(k^2)$ to play role to match the experiment data,
 except for $L_7$ and $L_8$. If
 the anomaly contribution donot play role, it must be cancelled in some sense
  and after the cancellation, we need to show the remaining dynamical effect
   (which may or may not be dominant by dynamical quark self energy) can still
    recover or improve the result (\ref{anomalyresult}). It is purpose of this
    work to judge these two possibilities.
     We will show the second choice is correct, the cancellation do happen in dynamical approach and remanent contribution
    from dynamical quark self energy $\Sigma(k^2)$ can provide values for
    parameters of chiral Lagrangian  consistent with experiment data.

 Consider QCD  in  presence of external scalar,pseudoscalar, vector and axial
  vector sources,
\begin{eqnarray}
J(x)=v\!\!\! /\;(x) +a\!\!\! /\;(x)\gamma_5 -s(x) +ip(x)\gamma_5\nonumber,
\end{eqnarray}
 The generating functional in Minkovski space is
\begin{eqnarray}
Z[J]&=&
\int{\cal D}\psi{\cal D}\overline{\psi}{\cal D}\Psi{\cal D}\overline{\Psi}
{\cal D}A_{\mu}\; e^{i{\int}d^{4}x[{\cal L}({\psi},\overline{\psi},\Psi,
\overline{\Psi},A_{\mu})+\overline{\psi}J\psi]}\label{genfund}
\end{eqnarray}
where $\psi,\Psi,A_{\mu}$ are light, heavy and gluon fields
respectively. ${\cal
L}({\psi},\overline{\psi},\Psi,\overline{\Psi},A_{\mu})$ is
Lagrangian of QCD. The chiral Lagrangian  relate this generating
functional by
\begin{eqnarray}
Z[J]&=&\int{\cal D}U\;\; e^{iS_{\rm GL}[U,J]}\label{genGS}
\end{eqnarray}
$U$ is pseudo goldstone boson (PGB) field, $S_{\rm GL}[U,J]$ is
 Gasser and Leutwyler (GL) chiral Lagrangian \cite{GS},
\begin{eqnarray}
S_{\rm GL}[U,J]&=&S_{\rm normal}[U,J]+S_{\rm anomaly}[U,J]\;,\\
S_{\rm normal}[U,J]&=&\int d^4x\bigg\{\frac{1}{4}F_0^2
{\rm tr}[\nabla^{\mu}U^{\dagger}\nabla_{\mu}U
+U\chi^{\dagger}+U^{\dagger}\chi]
+L_1[{\rm tr}(\nabla^{\mu}U^{\dagger}\nabla_{\mu}U)]^2\nonumber\\
&&+L_2{\rm tr}[\nabla_{\mu}U^{\dagger}\nabla_{\nu}U]
{\rm tr}[\nabla^{\mu}U^{\dagger}\nabla^{\nu}U]
+L_3{\rm tr}[(\nabla^{\mu}U^{\dagger}\nabla_{\mu}U)^2]\nonumber\\
&&+L_4{\rm tr}[\nabla^{\mu}U^{\dagger}\nabla_{\mu}U]
{\rm tr}[\chi^{\dagger}U+U^{\dagger}\chi]
+L_5{\rm tr}[\nabla^{\mu}U^{\dagger}\nabla_{\mu}U(\chi^{\dagger}U
+U^{\dagger}\chi)]\nonumber\\
&&+L_6[{\rm tr}(\chi^{\dagger}U+U^{\dagger}\chi)]^2
+L_7[{\rm tr}(\chi^{\dagger}U-U^{\dagger}\chi)]^2
+L_8{\rm tr}[\chi^{\dagger}U\chi^{\dagger}U
+\chi U^{\dagger}\chi U^{\dagger}]\nonumber\\
&&-iL_9{\rm tr}[F_{\mu\nu}^R\nabla^{\mu}U\nabla^{\nu}U^{\dagger}
+F_{\mu\nu}^L\nabla^{\mu}U^{\dagger}\nabla^{\nu}U]
+L_{10}{\rm tr}[U^{\dagger}F_{\mu\nu}^RUF^{L,\mu\nu}]\nonumber\\
&&+H_1{\rm tr}[F_{\mu\nu}^RF^{R,\mu\nu}+F_{\mu\nu}^LF^{L,\mu\nu}]
+H_2{\rm tr}[\chi^{\dagger}\chi]\bigg\}+O(p^6)\mbox{ terms}\label{GLdef}\\
S_{\rm anomaly}[U,J]&=&S_{\rm WZW}[U,J]+ O(p^6)\mbox{ terms}\;,
\end{eqnarray}
where
\begin{eqnarray}
\chi(x)=2B_0[s(x)+ip(x)]\label{chidef}
\end{eqnarray}
and $S_{\rm WZW}[U,J]$ is
Wess-Zumino-Witten action given in Ref.\cite{WZW}. Up to order of
$p^4$, $S_{\rm anomaly}[U,J]$ is completely known, but $S_{\rm
normal}[U,J]$ left fourteen parameters
$F_0,B_0,L_1,\ldots,L_{10},H_1,H_2$ need to be calculated.
To reveal the source of these parameters, we improve the conventional dynamical approach by building up a gauge invariant, nonlocal, dynamical (GND) quark
model. The action in GND model is assumed to be
$S_{\rm eff}[{\psi},\overline{\psi},U,J]$, it relate to our generating
functional by
\begin{eqnarray}
Z[J]&=&\int{\cal D}U{\cal D}\psi{\cal D}\overline{\psi}\;\;
e^{iS_{\rm eff}[{\psi},\overline{\psi},U,J]}\label{genGND}\;.
\end{eqnarray}
The r.h.s. of above equation can be seen as a result of integrating out heavy
quark and gluon fields and integrate in the PGB field $U$ in (\ref{genfund}).
If we further integrate out light quark field in above generating functional,
we obtain GL result (\ref{genGS}). So
$S_{\rm eff}[{\psi},\overline{\psi},U,J]$ can be seen as an intermediate stage
 action to relate fundamental QCD with phenomenological chiral Lagrangian.

Compare (\ref{genGS}) and (\ref{genGND}), we find GL chiral
Lagrangian relate to GND model by
\begin{eqnarray}
e^{iS_{\rm GL}[U,J]}=\int{\cal D}\psi{\cal D}\overline{\psi}\;\;
e^{iS_{\rm eff}[{\psi},\overline{\psi},U,J]}\label{Seffdef}\;,
\end{eqnarray}
$S_{\rm eff}[{\psi},\overline{\psi},U,J]$ is required to
 be invariant under following local $U_L(3)\otimes U_R(3)$
chiral transformations:
\begin{eqnarray}
&&\psi(x)\rightarrow\psi'(x)=[V_R(x)P_R+V_L(x)P_L]\psi(x)\nonumber\\
&&J(x)\rightarrow J'(x)=[V_R(x)P_L+V_L(x)P_R]
[J(x)+i\partial\!\!\!\! /\;]
[V_R^{\dagger}(x)P_R+V_L^{\dagger}(x)P_L]\nonumber\\
&&U(x)\rightarrow U'(x)=V_R(x)U(x)V_L^{\dagger}(x)\label{Utrans}\;.
\end{eqnarray}
Notice that $U$ field has standard decomposition
$U(x)=\Omega(x)\Omega(x)$ and $\Omega(x)$ field,
under transformation (\ref{Utrans}), transform as
$\Omega(x)\rightarrow\Omega'(x)
=h^{\dagger}(x)\Omega(x)V_L^{\dagger}(x)=V_R(x)\Omega(x)h(x)$
with $h(x)$ depend on $V_R,V_L$ and $\Omega$, represent an
induced hidden local $U(3)$ symmetry to keep transformed
$\Omega$ be a representative element at coset class.

To implement  local chiral symmetry explicitly, we take a special
local chiral transformation
$V_R(x)=\Omega^{\dagger}(x),V_L(x)=\Omega(x)$,
 the corresponding hidden symmetry transformation is $h(x)=1$,
\begin{eqnarray}
\psi_{\Omega}(x)&=&[\Omega^{\dagger}(x)P_R+\Omega(x)P_L]\psi(x)\label{psiOmega}
\\
J_{\Omega}(x)&=&[\Omega(x)P_R+\Omega^{\dagger}(x)P_L]
~[J(x)+i\partial\!\!\!\! /\;]
~[\Omega(x)P_R+\Omega^{\dagger}(x)P_L]\nonumber\\
&\equiv&-s_{\Omega}(x) +ip_{\Omega}(x)\gamma_5 +v\!\!\! /\;_{\Omega}(x)
+a\!\!\! /\;_{\Omega}(x)\gamma_5
\label{JOmegadef}\\
U_{\Omega}(x)&=&1\nonumber\;.
\end{eqnarray}
On rotated basis, we can rewrite (\ref{Seffdef}) as
\begin{eqnarray}
e^{iS_{GL}[U,J]}
&=&\frac{\int{\cal D}\psi{\cal D}\overline{\psi}\;\;
e^{iS_{\rm eff}[{\psi}_{\Omega},\overline{\psi}_{\Omega},1,J_{\Omega}]}}
{\int{\cal D}\psi{\cal D}\overline{\psi}e^{i\int d^4x
\overline{\psi}(x)[i\partial\!\!\!\! /\;_x+J(x)]\psi(x)}}~~
\int{\cal D}\psi{\cal D}\overline{\psi}e^{i\int d^4x
\overline{\psi}(x)[i\partial\!\!\!\! /\;_x+J(x)]\psi(x)}\nonumber\\
&=&N'\frac{\int{\cal D}\psi_{\Omega}{\cal D}\overline{\psi}_{\Omega}\;\;
e^{iS_{\rm eff}[{\psi}_{\Omega},\overline{\psi}_{\Omega},1,J_{\Omega}]}}
{\int{\cal D}\psi_{\Omega}{\cal D}\overline{\psi}_{\Omega}e^{i\int d^4x
\overline{\psi}_{\Omega}(x)[i\partial\!\!\!\! /\;_x+J_{\Omega}(x)]
\psi_{\Omega}(x)}}\label{GLdef1}
\end{eqnarray}
where $N'\equiv\int{\cal D}\psi{\cal D}\overline{\psi}e^{i\int d^4x
\overline{\psi}(x)[i\partial\!\!\!\! /\;_x+J(x)]\psi(x)}=
Det[i\partial\!\!\!\! /\;_x+J(x)]$.
In the last equality, we have taken chiral rotation (\ref{psiOmega}) for
 functional integration measure both in numerator and denominator.The possible
 anomalies caused by this rotation are cancelled between numerator and
  denominator.  Since we are only interested in $U$ dependence of
   the theory, pure source terms  $N'$ is irrelevant and therefore can
 be treated as a normalization factor.

Result (\ref{GLdef1}) tells us that $S_{\rm eff}$ should has following
structure
\begin{eqnarray}
S_{\rm eff}[{\psi}_{\Omega},\overline{\psi}_{\Omega},1,J_{\Omega}]
=\int d^4x\overline{\psi}_{\Omega}(x)[i\partial\!\!\!\! /\;_x+J_{\Omega}(x)]
\psi_{\Omega}(x)+
S_{\rm int}[{\psi}_{\Omega},\overline{\psi}_{\Omega},1,J_{\Omega}]
\end{eqnarray}
where $S_{\rm int}$ is interaction part caused by color gauge
interaction. If we switch off color gauge interaction which means
we are dealing with free fermion fields, there will be no
effective Lagrangian ($S_{\rm GL}=0$). $S_{\rm int}$ should
include those fermion self interaction terms caused by integrate
out gluon and heavy quark fields in underlying QCD and integrate in local
goldstone boson fields
 $U$. Among these, the most important effect related to chiral symmetry at low
energy region is spontaneous chiral symmetry breaking (SCSB) which
require quark has a nontrivial momentum dependent self energy
$\Sigma(k^2)$, its effects can be introduced into the theory by
adding in $S_{\rm int}$ a self energy term
\begin{eqnarray}
-\int
d^4x~\overline{\psi}_{\Omega}(x)\Sigma(\partial_x^2)\psi_{\Omega}(x)
\label{selfenergy}\;.
\end{eqnarray}
Just this term itself is not enough, since it is not invariant
under local chiral symmetry transformations. To make it invariant,
in conventional dynamical approach \cite{holdom}, a non-integratable face factor
is introduced into theory which cause very complex formulae and authors in
\cite{holdom} even donot put their analytical result in their papers. We donot use non-integratable face factor, instead we note that local chiral symmetry
transformation on rotated variable  is
\begin{eqnarray}
\psi_{\Omega}(x)&\rightarrow& \psi_{\Omega}'(x)=h^{\dagger}(x)\psi_{\Omega}(x)
\nonumber\\
J_{\Omega}(x)&\rightarrow&J_{\Omega}'(x)=h^{\dagger}(x)
[J_{\Omega}(x)+i\partial\!\!\!\! /\;_x]h(x)\label{hidden}\;.
\end{eqnarray}
 Original local chiral symmetry now is realized as  a hidden local symmetry.
Once the theory is constructed to be invariant under this hidden
symmetry, it is invariant under original local chiral symmetry.
Since interaction part in $S_{\rm eff}$ should be invariant
on local chiral symmetry, we need at least to generalize self energy
term (\ref{selfenergy}) to be invariant on hidden local  symmetry
(\ref{hidden}).
To achieve this, we change
the ordinary derivative $\partial^{\mu}_x$ to hidden symmetry
covariant derivative
$\overline{\nabla}^{\mu}_x=\partial^{\mu}_x-iv_{\Omega}^{\mu}(x)$
 (the overline on $\nabla^{\mu}_x$ is to denote the difference with
covariant derivative appeared in (\ref{GLdef})). (\ref{hidden})
tells us $v_{\Omega}(x)$ transform as
$v_{\Omega}^{\mu}(x)\rightarrow v_{\Omega}^{\mu\prime}(x)=h^{\dagger}(x)
v_{\Omega}^{\mu}(x)h(x)+ih^{\dagger}(x)[\partial^{\mu}h(x)]$
which lead
$\overline{\nabla}_x^{\mu}\rightarrow
\overline{\nabla}_x^{\mu\prime}=h^{\dagger}(x)\overline{\nabla}_x^{\mu}h(x)$.
The modified chiral invariant interaction action now is
\begin{eqnarray}
S_{\rm int}[{\psi}_{\Omega},\overline{\psi}_{\Omega},1,J_{\Omega}]
=-\int
d^4x~\overline{\psi}_{\Omega}(x)\Sigma(\overline{\nabla}_x^2)\psi_{\Omega}(x)
\label{Sint}\;.
\end{eqnarray}
This action is not the complete part of interaction, but it is the minimal
part of interaction which respect local chiral symmetry with dynamical quark
and SCSB. If we take the idea of dynamical perturbation originally from
Pagel-Stokar \cite{PS} and developed in Ref.\cite{HoldomWq}, in which at the
leading order of the expansion, all perturbative effects are ignored and only
nonperturbative effect considered in the theory is that from quark self
energy $\Sigma(k^2)$. (\ref{Sint}) in this sense  can be seen as a result of
leading order expansion from dynamical perturbation.

In GND model, quark fields dependence is bilinear and can be
exactly integrated out, the result GL Lagrangian from
(\ref{GLdef1}) is
\begin{eqnarray}
S_{GL}[U,J]&\approx& S_{GND}[U,J]\equiv -i{\rm
Tr}~ln[i\partial\!\!\!\!
/\;+J_{\Omega}-\Sigma(\overline{\nabla}^2)] +i{\rm
Tr}~ln[i\partial\!\!\!\! /\;+J_{\Omega}]\label{SGL}\;.
\end{eqnarray}

Use the Schwinger proper time formulation developed
in \cite{proper}, we can
compute the $\Sigma(\overline{\nabla}^2)$ dependent
determinant in (\ref{SGL}).  The result is
\begin{eqnarray}
&&-i{\rm Tr}~ln[i\partial\!\!\!\!
/\;+J_{\Omega}-\Sigma(\overline{\nabla}^2)]\label{p4}\\
&&=\int
d^4x{\rm tr}_f\bigg[B_0F_0^2s_{\Omega}+{\cal C}_1a_{\Omega}^2
+{\cal C}_2[d_{\mu}a_{\Omega}^{\mu}]^2
+{\cal C}_3(d^{\mu}a_{\Omega}^{\nu}-d^{\nu}a_{\Omega}^{\mu})
(d_{\mu}a_{\Omega,\nu}-d_{\nu}a_{\Omega,\mu}) \nonumber\\
&&\hspace{0.5cm}+{\cal C}_4[a_{\Omega}^2]^2
+{\cal C}_5a_{\Omega}^{\mu}a_{\Omega}^{\nu}a_{\Omega,\mu}a_{\Omega,\nu}
+{\cal C}_6s_{\Omega}^2+{\cal C}_7p_{\Omega}^2
+{\cal C}_8s_{\Omega}a_{\Omega}^2
 +{\cal C}_9V_{\Omega}^{\mu\nu}V_{\Omega,\mu\nu}
 +{\cal C}_{10}V_{\Omega}^{\mu\nu}a_{\Omega,\mu}a_{\Omega,\nu}
\nonumber\\ &&\hspace{0.5cm}
 +{\cal C}_{11}p_{\Omega}d_{\mu}a^{\mu}_{\Omega}\bigg] +O(p^6)
 +\mbox{\rm imaginary terms} \nonumber
\end{eqnarray}
 where tr$_f$ is trace for flavor indices. Covariant derivative for function
 $f$ and $V^{\mu\nu}_{\Omega}$ are defined as
\begin{eqnarray}
d^{\mu}f\equiv\partial^{\mu}f-iv^{\mu}_{\Omega}f+ifv^{\mu}_{\Omega}
\label{dmudef}\hspace{2cm}
V_{\Omega}^{\mu\nu}=\partial^{\mu}v_{\Omega}^{\nu}
-\partial^{\nu}v_{\Omega}^{\mu}
-iv_{\Omega}^{\mu}v_{\Omega}^{\nu}+iv_{\Omega}^{\nu}v_{\Omega}^{\mu}\,,
\;.
\end{eqnarray}
   The $\Sigma$ dependence for coefficients appeared in (\ref{p4}) are
\begin{eqnarray}
&&F_0^2B_0=4\int d\tilde{k} \Sigma_kX_k\label{F0B0}\\
&&{\cal C}_1=2\int d\tilde{k}\bigg[
(-2\Sigma^2_k-k^2\Sigma_k\Sigma'_k)X_k^2+(2\Sigma^2_k+k^2\Sigma_k\Sigma'_k)
\frac{X_k}{\Lambda^2}\bigg]\label{F0}\\
&&{\cal C}_2=-2\int d\tilde{k}\bigg[-2A_kX_k^3
+2A_k\frac{X_k^2}{\Lambda^2}-A_k\frac{X_k}{\Lambda^4}
+\frac{k^2}{2}\Sigma'^2_k\frac{X_k}{\Lambda^2}-\frac{k^2}{2}\Sigma'^2_kX_k^2
\bigg]\nonumber\\
&&{\cal C}_3=-\int d\tilde{k}\bigg[-2B_kX_k^3
 +2B_k\frac{X_k^2}{\Lambda^2}-B_k\frac{X_k}{\Lambda^4}
+\frac{k^2}{2}\Sigma^{\prime 2}_k\frac{X_k}{\Lambda^2}
-\frac{k^2}{2}\Sigma^{\prime 2}_kX_k^2\bigg]\nonumber\\
&&{\cal C}_4=2\int d\tilde{k}\bigg[
(\frac{4\Sigma^4_k}{3}-\frac{2k^2\Sigma^2_k}{3}+\frac{k^4}{18})(
 6X_k^4-\frac{6X_k^3}{\Lambda^2}+\frac{3X_k^2}{\Lambda^4}
-\frac{X_k}{\Lambda^6})+(-4\Sigma^2_k+\frac{k^2}{2})(
-2X_k^3\nonumber\\
&&\hspace{1.5cm}+\frac{2X_k^2}{\Lambda^2}-\frac{X_k}{\Lambda^4})
-\frac{X_k}{\Lambda^2}+X_k^2\bigg]\nonumber\\
&&{\cal C}_5=\int d\tilde{k}\bigg[
(\frac{-4\Sigma^4_k}{3}+\frac{2k^2\Sigma^2_k}{3}+\frac{k^4}{18})(
6X_k^4-\frac{6X_k^3}{\Lambda^2}+\frac{3X_k^2}{\Lambda^4}
-\frac{X_k}{\Lambda^6})
+4\Sigma^2_k(-2X_k^3+\frac{2X_k^2}{\Lambda^2}\nonumber\\
&&\hspace{1.5cm}
-\frac{X_k}{\Lambda^4})+\frac{X_k}{\Lambda^2}-X_k^2\bigg]\nonumber\\
&&{\cal C}_6=2\int d\tilde{k}\bigg[(3\Sigma^2_k+2k^2\Sigma_k\Sigma'_k)X_k^2
+[-2\Sigma^2_k-k^2(1+2\Sigma_k\Sigma'_k)]\frac{X_k}{\Lambda^2}\bigg]\nonumber\\
&&{\cal C}_7=2\int d\tilde{k}\bigg[(\Sigma^2_k+2k^2\Sigma_k\Sigma'_k)X_k^2
-k^2(1+2\Sigma_k\Sigma'_k)\frac{X_k}{\Lambda^2}\bigg]\nonumber\\
&&{\cal C}_8=4\int d\tilde{k}\bigg[(-4\Sigma^3_k+k^2\Sigma_k)X_k^3
+(4\Sigma^3_k-k^2\Sigma_k)\frac{X_k^2}{\Lambda^2}
-(2\Sigma^3_k-\frac{1}{2}k^2\Sigma_k)\frac{X_k}{\Lambda^4}
+3\Sigma_k\frac{X_k}{\Lambda^2}
-3\Sigma_kX_k^2\bigg]\nonumber\\
&&{\cal C}_9=-\int d\tilde{k}\bigg[
(\frac{1}{3}k^2\Sigma'_k\Sigma''_k+\frac{1}{3}\Sigma_k\Sigma''_k)X_k
+(C_k-D_k)\frac{X_k}{\Lambda^2}
-(C_k-D_k)X_k^2-2E_kX_k^3\nonumber\\
&&\hspace{1.5cm}+2E_k\frac{X_k^2}{\Lambda^2}
-E_k\frac{X_k}{\Lambda^4}\bigg]\nonumber\\
&&i{\cal C}_{10}=4\int d\tilde{k}\bigg[
-2F_kX_k^3+2F_k\frac{X_k^2}{\Lambda^2}
-F_k\frac{X_k}{\Lambda^4}
+\frac{k^2}{2}\Sigma_k^{\prime 2}\frac{X_k}{\Lambda^2}
-\frac{k^2}{2}\Sigma^{\prime 2}_kX_k^2\bigg]
 \nonumber\\
&&{\cal C}_{11}=-4\int d\tilde{k}\bigg[
-(\Sigma_k+\frac{1}{2}k^2\Sigma'_k)\frac{X_k}{\Lambda^2}
+(\Sigma_k+\frac{1}{2}k^2\Sigma'_k)X_k^2\bigg]\label{Kresult}
\end{eqnarray}
where
\begin{eqnarray}
\int d\tilde{k}&\equiv&
iN_c\int\frac{d^4k}{(2\pi)^4}e^{\frac{k^2-\Sigma^2(-k^2)}{\Lambda^2}}
\label{measure}\hspace{2cm}
\Sigma_k\equiv\Sigma(-k^2)\hspace{2cm}
X_k\equiv\frac{1}{k^2-\Sigma^2(-k^2)}\nonumber\\
A_k&=&\frac{2}{3}k^2\Sigma_k\Sigma'_k(1+2\Sigma_k\Sigma'_k)+\frac{1}{3}
\Sigma^2_k(1+2\Sigma_k\Sigma'_k)
-\frac{1}{3}k^2\Sigma^2_k(\Sigma^{\prime 2}_k+\Sigma_k\Sigma''_k)
+\frac{1}{6}k^4(\Sigma^{\prime 2}_k+\Sigma_k\Sigma''_k) \nonumber\\
B_k&=&\frac{2}{3}k^2\Sigma_k\Sigma'_k(1+2\Sigma_k\Sigma'_k)+\frac{1}{3}
\Sigma^2_k(1+2\Sigma_k\Sigma'_k)-\frac{1}{3}k^2\Sigma^2_k(\Sigma^{\prime
2}_k +\Sigma_k\Sigma''_k)
+\frac{1}{18}k^4(\Sigma^{\prime 2}_k+\Sigma_k\Sigma''_k)
\nonumber\\
&&+\frac{1}{6}k^2(1+2\Sigma_k\Sigma'_k)\nonumber\\
C_k&=&\frac{1}{3}-\frac{1}{3}\Sigma_k\Sigma'_k
-\frac{1}{2}k^2\Sigma^{\prime 2}_k\nonumber\\
D_k&=&\frac{1}{2}k^2\Sigma^{\prime
2}_k+\frac{1}{6}k^2\Sigma_k\Sigma''_k
(1+2\Sigma_k\Sigma'_k)+\frac{2}{9}k^4\Sigma'_k\Sigma''_k
(1+2\Sigma_k\Sigma'_k) +\frac{2}{9}k^4\Sigma^{\prime
2}_k(\Sigma^{\prime 2}_k+\Sigma_k\Sigma''_k)\nonumber\\
&&+\frac{1}{3}k^2\Sigma_k\Sigma'_k(\Sigma^{\prime
2}_k+\Sigma_k\Sigma''_k) \nonumber\\
E_k&=&-\frac{1}{6}k^2\Sigma_k\Sigma'_k(1+2\Sigma_k\Sigma'_k)^2
-\frac{1}{9}k^4\Sigma^{\prime
2}_k(1+2\Sigma_k\Sigma'_k)^2\nonumber\\
F_k&=&-\frac{4}{3}k^2\Sigma_k\Sigma'_k+
\frac{4}{3}k^2(\Sigma_k\Sigma'_k)^2-\frac{2}{3}\Sigma^2_k
+\frac{2}{3}\Sigma^3_k\Sigma'_k
-\frac{1}{3}k^2\Sigma^2_k(\Sigma^{\prime
2}_k+\Sigma_k\Sigma''_k)
+\frac{1}{9}k^4(\Sigma^{\prime 2}_k+\Sigma_k\Sigma''_k)
\nonumber\\
&&+\frac{1}{3}k^2(1+2\Sigma_k\Sigma'_k)-\frac{1}{2}k^2\nonumber\;.
\end{eqnarray}
The result for ${\cal C}_1\equiv F_0^2$ in (\ref{F0}) is just the well
known Pagel-Stokar formula \cite{PS}, if we take momentum cutoff
$\Lambda$ be infinity. The part of $\Sigma(\overline{\nabla}^2)$
independent quark determinant in (\ref{p4}) is just the result
of anomaly approach with a total minus sign, we can get  result of anomaly
approach by taking limit of $\Sigma_k={\rm const}\rightarrow 0$
 (Note due to possible infrared divergence, limit of $\Sigma\rightarrow
0$ must be taken after the momentum integration). The nonzero
coefficients ${\cal C}_i$ for the case of infinite momentum cutoff $\Lambda$ is
\begin{eqnarray}
&&{\cal C}_2 \stackrel{\Sigma\rightarrow 0}{--\rightarrow}
\frac{N_c}{24\pi^2}\;,~~~~~
{\cal C}_3\stackrel{\Sigma\rightarrow 0}{--\rightarrow}
\frac{N_c}{48\pi^2}(ln\frac{\Sigma^2}{\Lambda^2}+\gamma+1)\;,~~~~~
{\cal C}_4\stackrel{\Sigma\rightarrow 0}{--\rightarrow}
\frac{N_c}{24\pi^2}(ln\frac{\Sigma^2}{\Lambda^2}+\gamma+4)\;,\nonumber\\
&&{\cal C}_5\stackrel{\Sigma\rightarrow 0}{--\rightarrow}
-\frac{N_c}{24\pi^2}(ln\frac{\Sigma^2}{\Lambda^2}+\gamma+2)\;,~~~~~
{\cal C}_6\stackrel{\Sigma\rightarrow 0}{--\rightarrow}
\frac{N_c}{8\pi^2}\Lambda^2\;,~~~~~
{\cal C}_7\stackrel{\Sigma\rightarrow 0}{--\rightarrow}
\frac{N_c}{8\pi^2}\Lambda^2 \nonumber\\
&&{\cal C}_9\stackrel{\Sigma\rightarrow 0}{--\rightarrow}
\frac{N_c}{48\pi^2}(ln\frac{\Sigma^2}
{\Lambda^2}+\gamma)\;,~~~~~
i{\cal C}_{10}\stackrel{\Sigma\rightarrow 0}{--\rightarrow}
\frac{N_c}{12\pi^2}(ln\frac{\Sigma^2}{\Lambda^2}+\gamma+2)\nonumber\;\;.
\end{eqnarray}

 The pure imaginary  terms in (\ref{p4}) are completely
  known at phenomenological level and its calculation in terms of $\Sigma$
  is already performed in Ref.\cite{GCM} and proved exactly recover the
  Witten's result \cite{WZW}, we donot explicitly write down their detail
  structures.

With help of (\ref{JOmegadef}), (\ref{SGL}) and (\ref{p4}) lead relation,
\begin{eqnarray}
&&L_1=\frac{1}{2}L_2=\frac{\overline{\cal C}_5}{32}
-\frac{\overline{\cal C}_9}{16}
+i\frac{\overline{\cal C}_{10}}{32},~~~
L_3=\frac{\overline{\cal C}_4-2\overline{\cal C}_5+6\overline{\cal C}_9
-3i\overline{\cal C}_{10}}{16},\nonumber\\
&&L_4=0,~~~
L_5=\frac{\overline{\cal C}_8}{16B_0},
~~~L_6=0,~~~L_7=\frac{\overline{\cal C}_2}{48}
-\frac{\overline{\cal C}_{11}}{48B_0},\nonumber\\
&&L_8=-\frac{\overline{\cal C}_2}{16}
+\frac{\overline{\cal C}_6}{16B_0^2}
-\frac{\overline{\cal C}_7}{16B_0^2}
+\frac{\overline{\cal C}_{11}}{16B_0},~~~
L_9= \frac{-4\overline{\cal C}_9
+i\overline{\cal C}_{10}}{8},\nonumber\\
&&L_{10}=\frac{-\overline{\cal C}_3 +\overline{\cal C}_9}{2},~
H_1=\frac{\overline{\cal C}_3 +\overline{\cal C}_9}{4},~~~
H_2=\frac{\overline{\cal C}_2}{8} +\frac{\overline{\cal
C}_6}{8B_0^2} +\frac{\overline{\cal C}_7}{8B_0^2}
-\frac{\overline{\cal C}_{11}}{8B_0}.\label{p4a}
\end{eqnarray}
where $\overline{\cal C}_i\equiv{\cal C}_i
-\lim_{\Sigma\rightarrow 0}{\cal C}_i\hspace{0.5cm}i=1,2,\ldots,11$.
Term $-\lim_{\Sigma\rightarrow 0}{\cal C}_i$ is of special interest, since it
 relate to anomaly result mentioned in the
beginning of this paper. In fact, in anomaly approach, the
effective action is $i{\rm
Tr}~ln[i\partial\!\!\!\!/\;+J_{\Omega}]$ \cite{anomalyderive},
which is just the result of (\ref{GLdef}) with (\ref{p4a}) by taking
 ${\cal C}_i$ values at $\Sigma=0$ and revert
all signs. One can easily check this reproduce result
(\ref{anomalyresult}) in which all ultraviolet divergence are
cancelled each
 other  for $L_i$ parameters. The interpretation of this result is that
 from (\ref{SGL}), {\it the contribution of anomaly play
no role in the final result , it is completely cancelled by
dynamical quark self energy dependent part,
only the remainder after cancellation play
 role in the final parameters $L_i$.}

In (\ref{p4a}), parameter $B_0$ needs special treatment. Since with help of
 (\ref{F0B0}), we find $F_0^2B_0$ is generally divergent.
To renormalize this condensate, we note that
$F_0$ and $m_{\Lambda}\langle\overline{\psi}\psi\rangle_{\Lambda}$
($m_{\Lambda}$ is bare current quark mass) is renormalization invariant or
more general, the $\chi$ field defined in (\ref{chidef}) is renormalization
invariant, i.e.  $\chi(x)=2B_0[s(x)+ip(x)]=2B_r[s_r(x)+ip_r(x)]$
with $B_r,s_r,p_r$ are renormalized $B_0,s,p$. Correspondingly,
renormalized quark condensate
$\langle\overline{\psi}\psi\rangle_r$ is
$\langle\overline{\psi}\psi\rangle_r=-N_fF_0^2B_r$.
So replacing the scalar and peudoscalar sources and
$B_0$ with renormalized ones donot change value of $\chi$ field.
With renormalized sources, we can replace $B_0$ in (\ref{p4a}) with $B_r$.
In this paper, we donot directly calculate and use $B_0$,
instead we calculate and use $B_r$. The renormalization point is chosen to be
at scale of 1GeV.

Now, once  the quark self energy $\Sigma(k^2)$ was input into
the formulae, we can get all parameters in Gasser-Leutwyler
chiral Lagrangian. In conventional dynamical approach \cite{holdom},
 the ignorance of $\Sigma(k^2)$ is  parametrized by following ansatz
\begin{eqnarray}
\Sigma(k^2)=\frac{(A+1)m^3}{k^2+Am^2}\label{ansatz}
\end{eqnarray}
which satisfy $\Sigma(m^2)=m$ and shares qualitative similarities
with solutions of improved ladder SD equation. It is finite,
positive, monotonically decreasing functions with $1/k^2$ behavior
at large $q^2$ and $\Sigma'(0)<0$. The constituent quark mass $m$
is determined for each choice of $A$ for $F_0=93MeV$ from
(\ref{F0}). For $A=1,2,3,4$, we obtain $m=379,350,331,317MeV$
respectively.  With ansatz (\ref{ansatz}), the result parameters are listed
in TABLE.\ref{I}. We see that the wrong sign problem for $L_7$ and $L_8$
in anomaly calculation  is corrected now and result parameters are roughly
consistent with experiment data.

To further trace  the relation of GND model with underlying theory QCD.
Note that the parameters in the chiral Lagrangian
are recently expressed in terms of QCD Green's
functions \cite{WQ} and for quark two point Green's function
$\Phi^{\sigma\rho}_{\Omega}(x,y)$, at large $N_c$ limit, \cite{WQ} gives
equation,
\begin{eqnarray}                            
&&[i\partial\!\!\! /\;+i\Phi_{\Omega}^{T,-1}+J_{\Omega}
+\tilde{\Xi}]^{\sigma\rho}(x,y)
+\sum^{\infty}_{n=1}{\int}d^{4}x_1\cdots{d^4}x_{n}
d^{4}x_{1}'\cdots{d^4}x_{n}'\frac{(-i)^{n+1}(N_c g^2)^n}{n!}\nonumber\\
&&\times\overline{G}^{\sigma\sigma_1\cdots\sigma_n}_{\rho\rho_1
\cdots\rho_n}(x,y,x_1,x'_1,\cdots,x_n,x'_n)
\Phi_{\Omega}^{\sigma_1\rho_1}(x_1 ,x'_1)\cdots
\Phi_{\Omega}^{\sigma_n\rho_n}(x_n ,x'_n)=0.\label{fineq}
\end{eqnarray}

Now consider the coincidence limit of two point quark Green's
function in QCD in presence of external sources,
\begin{eqnarray}
&&\frac{1}{N_c}\langle 0|{\bf\rm T}
\overline{\psi}_{\alpha}^{a\eta}(x)\psi_{\alpha}^{b\xi}(x')
|0\rangle_{\rm QCD}
\equiv\frac{-i}{N_c}\frac{\delta \ln Z[J]}{\delta J^{(a\eta)(b\xi)}(x)}
\nonumber\\
&&=\frac{1}{N_c}\frac{
\int{\cal D}U\;\;
\frac{\delta S_{\rm GL}[U,J]}{\delta J^{(a\eta)(b\xi)}(x)}
e^{iS_{\rm GL}[U,J]}}
{\int{\cal D}U\;\; e^{iS_{\rm GL}[U,J]}}
=\frac{\int{\cal D}U\;\;\Phi^{(a\eta)(b\xi)}[U,J](x,x)~
e^{iS_{\rm GL}[U,J]}}
{\int{\cal D}U\;\; e^{iS_{\rm GL}[U,J]}}
\end{eqnarray}
where we have used (\ref{genGS}) and $\Phi^{(a\eta)(b\xi)}[U,J](x,x)$
is
\begin{eqnarray}
\Phi^{(a\eta)(b\xi)}[U,J](x,x)&\equiv&
\frac{1}{N_c}
\frac{\delta S_{\rm GL}[U,J]}{\delta J^{(a\eta)(b\xi)}(x)}
\approx\frac{1}{N_c}
\frac{\delta S_{\rm GND}[U,J]}{\delta J^{(a\eta)(b\xi)}(x)}\;.
\end{eqnarray}
Use (\ref{SGL}), the rotated $\Phi$ become
\begin{eqnarray}
\Phi^{(a\eta)(b\xi)}_{\Omega}[U,J](x,x) &=&{\rm
Tr}\bigg[\frac{-i}{i\partial\!\!\!\! /\;+J_{\Omega}
-\Sigma(\overline{\nabla}^2)} \frac{\delta
[J_{\Omega}-\Sigma(\overline{\nabla}^2)]} {\delta
J^{(a\eta)(b\xi)}_{\Omega}(x)}\bigg] \label{Phidef}
\end{eqnarray}
which imply
\begin{eqnarray}
\Phi_{\Omega}(x,y) &=&-i\int d^4z~[i\partial\!\!\!\!
/\;+J_{\Omega}
-\Sigma(\overline{\nabla}^2)]^{-1}(x,z)[\delta(z-y)+\Delta(z,y)]
\end{eqnarray}
where $\Delta(z,y)$ relate to
$\delta\Sigma(\overline{\nabla}^2)/\delta J$. Compare to (\ref{fineq}),
we find, present choice of $S_{\rm int}$ is equivalent to take
following approximation
\begin{eqnarray}
&&\bigg[(1+\Delta)^{-1}[\Sigma(\overline{\nabla}^2)
-\Delta(i\partial\!\!\! /\;+J_{\Omega})]\bigg]^{\sigma\rho}(x,y)\nonumber\\
&&\approx\tilde{\Xi}^{\sigma\rho}(x,y)
+\sum^{\infty}_{n=1}{\int}d^{4}x_1\cdots{d^4}x_{n}
d^{4}x_{1}'\cdots{d^4}x_{n}'\frac{(-i)^{n+1}(N_c g^2)^n}{n!}\nonumber\\
&&\hspace{0.5cm}\times\overline{G}^{\sigma\sigma_1\cdots\sigma_n}_{\rho\rho_1
\cdots\rho_n}(x,y,x_1,x'_1,\cdots,x_n,x'_n)
\Phi_{\Omega}^{\sigma_1\rho_1}(x_1 ,x'_1)\cdots
\Phi_{\Omega}^{\sigma_n\rho_n}(x_n ,x'_n)\label{DSeq}\;.
\end{eqnarray}
If we further  drop correlation functions
$\overline{G}^{\sigma\sigma_1\cdots\sigma_n}_{\rho\rho_1\cdots\rho_n}
(x,y,x_1,x'_1,\cdots,x_n,x'_n)$ with $n>1$
and ignore the external sources in above equation ($\Delta$
therefore must be ignored), as mentioned in Ref. \cite{WQ},
(\ref{DSeq}) then is just Schwinger-Dyson equation for quark
propagator. We only consider the term of quark
self energy with argument of $\overline{\nabla}^2$   which is the minimal
generalization from pure $\Sigma(\partial^2)$ term to source
dependent terms satisfying local chiral symmetry.
Include in source terms, just self energy term in
l.h.s. of (\ref{DSeq}) is not enough to match all contributions of
r.h.s. of equation, but if underlying QCD can provide correct predictions for
 parameters in chiral Lagrangian, the fact of $\Sigma(k^2)$ dominance in
 the parameters of chiral Lagrangian
 imply that the terms we dropped at this work should not play so
 important role. we will leave these terms in addition to self
energy term to balance equation (\ref{DSeq}) in future
investigations.

In conclusion, all 12 parameters in $p^4$ order $SU(3)$ chiral
Lagrangian are explicitly expressed in terms of functions of quark
self energy $\Sigma(k^2)$. We have shown that the original result given
from anomaly approach is completely cancelled. Take suitable
quark self energy ansatz to perform numerical calculation, we find
after cancellation of anomaly contribution,
 the dynamical quark self energy do can provide
parameter values roughly consistent with experimental data.

\section*{Acknowledgments}

This work was  supported by National  Science Foundation of China No.90103008
and fundamental research grant of Tsinghua University.


\begin{table}
\caption{\label{I}Values, multiplied by $10^3$, for the parameters of the
order $p^4$ chiral Lagrangian calculated in GND model with  quark self energy
determined by ansatz (\ref{ansatz}). anomaly: anomaly approach result;
expt: experimental values. }
\begin{tabular}{|c|c|c|c|c|c|c|c|c|c|c|}
&$L_1$ &$L_2$ & $L_3$ & $L_4$ & $L_5$ &
$L_6$ & $L_7$ & $L_8$ & $L_9$ & $L_{10}$
 \\ \hline\hline
A=1 &0.927&1.85&-6.55&0&1.79 &0&-0.570 &1.56&3.79&-5.18 \\ \hline
A=2 &0.771&1.54&-5.53&0&1.63 &0 &-0.501 &1.39&2.60&-3.91 \\ \hline
A=3 &0.708&1.42&-5.09&0&1.51 &0&-0.449 &1.26&2.13&-3.38\\ \hline
A=4 &0.674&1.35&-4.85&0&1.41 &0 &-0.413 &1.17&1.90&-3.08 \\
\hline\hline
anomaly  & 0.792 & 1.58 & -3.17 & 0 & 0 & 0 & 0.263 & -0.792
 & 6.33 & -3.17 \\ \hline
expt   & 0.9$\pm$0.3 & 1.7$\pm$0.7 & -4.4$\pm$2.5 & 0$\pm$0.5 &
2.2$\pm$0.5 & 0$\pm$0.3 & -0.4$\pm$0.15 & 1.1$\pm$0.3 & 7.4$\pm$0.7 &
-6.0$\pm$0.7 \\
\end{tabular}
\end{table}

\end{document}